\documentclass[12pt]{article}
\usepackage{graphicx}
\usepackage{amsfonts}
\usepackage{amsmath}
\usepackage{amssymb}
\usepackage{subfigure}
\usepackage{cite}
\usepackage{mathrsfs}
\usepackage{color}

\newcommand\pubdate{\today}

\def\cftp{Centro de F\'isica Te\'orica de Part\'iculas,\\
Instituto Superior T\'ecnico, Universidade de Lisboa,\\ Av. Rovisco Pais 1, 1049-001 Lisbon, PORTUGAL}
\def\support{\footnote{Work supported by \emph{Funda\c{c}\~ao para a Ci\^encia e a Tecnologia} (FCT, Portugal) through the projects CERN/FP/123580/2011 and CFTP-FCT Unit 777.}}

\def\Title#1{\begin{center} {\Large #1 } \end{center}}
\def\Author#1{\begin{center}{ \sc #1} \end{center}}
\def\Address#1{\begin{center}{ \it #1} \end{center}}

\newcommand\pubblock{\rightline{\begin{tabular}{l} \pubdate  \end{tabular}}}
\newenvironment{Abstract}{\begin{quotation}  }{\end{quotation}}
\newenvironment{Presented}{\begin{quotation} \begin{center} 
             PRESENTED AT\end{center}\bigskip 
      \begin{center}\begin{large}}{\end{large}\end{center} \end{quotation}}

\def\beq{\begin{equation}}
\def\eeq#1{\label{#1}\end{equation}}
\def\eeqn{\end{equation}}

\def\beqa{\begin{eqnarray}}
\def\eeqa#1{\label{#1}\end{eqnarray}}
\def\eeqan{\end{eqnarray}}

\let\bar=\overbar

\def\Dslash{\not{\hbox{\kern-4pt $D$}}}
\def\dslash{\not{\hbox{\kern-2pt $\del$}}}

\def\msb{{\bar{\ssstyle M \kern -1pt S}}}

\newcommand{\Mmix}[1]{M_{12}^{({#1})}}
\newcommand{\Mmixq}{\Mmix{q}}
\newcommand{\Mmixd}{\Mmix{d}}
\newcommand{\Mmixs}{\Mmix{s}}
\newcommand{\Gmix}[1]{\Gamma_{12}^{({#1})}}
\newcommand{\Gmixq}{\Gmix{q}}

\newcommand{\DMBq}{\Delta M_{B_q}}
\newcommand{\DC}{\Delta\chi^2}

\newcommand{\asld}{A^d_{SL}}
\newcommand{\asls}{A^s_{SL}}
\newcommand{\aslb}{A^b_{SL}}
\newcommand{\aslq}{A^q_{SL}}

\newcommand{\BBdmix}{$B^0_d$--$\bar B^0_d$}
\newcommand{\BBsmix}{$B^0_s$--$\bar B^0_s$}

\newcommand{\DGd}{\Delta\Gamma_d}
\newcommand{\DGs}{\Delta\Gamma_s}
\newcommand{\DGq}{\Delta\Gamma_q}

\newcommand{\AJPKs}{A_{J/\Psi K_S}}
\newcommand{\AJPP}{A_{J/\Psi \Phi}}

\newcommand{\CKM}{V}
\newcommand{\V}[1]{\CKM_{#1}^{\,}}
\newcommand{\Vc}[1]{\CKM_{#1}^{\ast}}
\newcommand{\La}[2]{\lambda^{#1}_{#2}}

\newcommand{\re}[1]{\text{Re}\left(#1\right)}
\newcommand{\im}[1]{\text{Im}\left(#1\right)}

\newcommand{\refeq}[1]{(\ref{#1})}
\newcommand{\eq}[1]{eq.\refeq{#1}}

\graphicspath{{Figs/}}

\begin{document}
\begin{titlepage}
\pubblock

\vfill
\Title{The like-sign dimuon asymmetry and New Physics}
\vfill
\Author{Miguel Nebot\support}
\Address{\cftp}
\vfill
\begin{Abstract}
\noindent The measurement by the D0 collaboration of a large like-sign dimuon asymmetry deviates significantly from Standard Model expectations. New Physics may be invoked to account for such a deviation. We analyse how generic extensions of the Standard Model where the Cabibbo-Kobayashi-Maskawa $3\times 3$ mixing matrix is enlarged can accommodate a significant enhancement of $\aslb$ with respect to standard expectations through enhancements of the individual semileptonic asymmetries $\asld$ and $\asls$ in the \BBdmix\ and \BBsmix\ systems. The potential enhancement reachable in this class of scenario is, nevertheless, insufficient to reproduce the D0 measurement.
\end{Abstract}
\vfill
\begin{Presented}
8th International Workshop on the CKM Unitarity Triangle (CKM 2014),\\ Vienna, Austria, September 8-12, 2014
\end{Presented}
\vfill
\end{titlepage}
\def\thefootnote{\fnsymbol{footnote}}
\setcounter{footnote}{0}

\section{Introduction}

Phenomena related to Flavour Physics and CP violation constitute a fundamental window to probe the Standard Model (SM) and its extensions. In this context, one results from the D0 collaboration has received much attention: the measurement of the like-sign dimuon asymmetry $\aslb$ \cite{Abazov:2013uma}. 
 The value reported by the D0 collaboration \cite{Abazov:2013uma} is approximately ``$3\sigma$'' away from SM expectations, and a large number of works have explored the potential of models beyond the SM to reproduce it \cite{Ko:2010mn}.
 In the following, we first review the SM predictions and then address NP analyses, focusing on scenarios where the mixing matrix is enlarged with respect to the usual $3\times 3$ unitary Cabibbo-Kobayashi-Maskawa. 

\section{Mixing and asymmetries in $\mathbf{B_q}$ meson systems\label{SEC:mix}}

In the SM, the dispersive $B_q\to\bar B_q$ transition amplitude, $\Mmixq$, is dominated by one loop box diagrams with virtual $t$ quarks:
\begin{equation}
\left[\Mmixq\right]_{\text{SM}}=\frac{G_F^2M_W^2}{12\pi^2}\,M_{B_q}\,f_{B_q}^2\,B_{B_q}\,\eta_B\,(\V{tb}\Vc{tq})^2\,S_0(x_t)\ .\label{eq:M12q:01}
\end{equation}
The absorptive part, $\Gmixq$, is on the contrary dominated by intermediate real (on-shell) $u$ and $c$ quarks. The SM short-distance prediction \cite{Beneke:1998sy} requires a Heavy Quark Expansion giving $\Gmixq$ as an expansion in $\alpha_s(m_b)$ and $\Lambda/m_b$. Our interest lies on the flavour structure, which has, in general, the following form
\begin{equation}
\frac{\Gmixq}{\Mmixq}= -\Bigg[\frac{\Gamma_{12}^{cc}}{\Mmixq}(\V{cb}\Vc{cq})^2
+2\frac{\Gamma_{12}^{uc}}{\Mmixq}(\V{ub}\Vc{uq}\V{cb}\Vc{cq}) +\frac{\Gamma_{12}^{uu}}{\Mmixq}(\V{ub}\Vc{uq})^2\Bigg]\,,\label{eq:G12q:01}
\end{equation}
and in particular in the SM the flavour structure is
\begin{equation}
\left[\frac{\Gmixq}{\Mmixq}\right]_{\textrm{SM}}\propto 
\Gamma_{12}^{cc}\frac{(\V{cb}\Vc{cq})^2}{(\V{tb}\Vc{tq})^2}+2\Gamma_{12}^{uc}\frac{\V{ub}\Vc{uq}\V{cb}\Vc{cq}}{(\V{tb}\Vc{tq})^2}+\Gamma_{12}^{uu}\frac{(\V{ub}\Vc{uq})^2}{(\V{tb}\Vc{tq})^2}
\ .\label{eq:G12q:01b}
\end{equation}
$\Gamma_{12}^{ab}$ are $-\Gamma_{12}^{cc}=c$, $-2\Gamma_{12}^{uc}=2c-a$, $-\Gamma_{12}^{uu}=b+c-a$, where
\begin{equation}
a= (10.5\pm 1.8)\cdot 10^{-4}\,,\ 
b= (0.2\pm 0.1)\cdot 10^{-4}\,,\ 
c= (-53.3\pm 12.0)\cdot 10^{-4}\,.\label{eq:G12d:01:coeff2}
\end{equation}
In an expansion in powers of $(m_c/m_b)^2$, it is important to stress that at zero-th order only $c$ is present.
Unitarity of the CKM mixing matrix, through the orthogonality condition $\V{ub}\Vc{uq}+\V{cb}\Vc{cq}+\V{tb}\Vc{tq}=0$, implies
\begin{equation}
\left[\frac{\Gmixq}{\Mmixq}\right]_{\textrm{SM}}= 
 K_{(q)}\left[c+a\,\frac{\V{ub}\Vc{uq}}{\V{tb}\Vc{tq}}+b\,\left(\frac{\V{ub}\Vc{uq}}{\V{tb}\Vc{tq}}\right)^2\right]\ ,\label{eq:G12q:03}
\end{equation}
with $K_{(q)}=\frac{12\pi^2}{M_{B_q}\,f_{B_q}^2\,B_{B_q}\,G_F^2M_W^2\,\eta_B\,S_0(x_t)}$.\\ 
$\Gmixq/\Mmixq$ is accessed through the width difference $\DGq$ and the genuinely CP violating semileptonic asymmetry $\aslq$; at leading order in $\Gmixq/\Mmixq$,
\begin{equation}
-\frac{\DGq}{\DMBq}=\re{\frac{\Gmixq}{\Mmixq}}\,,\quad \aslq=\im{\frac{\Gmixq}{\Mmixq}} \,.\label{eq:G12qObs:01}
\end{equation}
The SM expectations are
\begin{eqnarray}
&\left[\asld\right]_{\textrm{SM}}  = (-4.2\pm 0.7)\cdot 10^{-4}\,,\nonumber\quad
&\left[\DGd\right]_{\textrm{SM}}  = (2.60\pm 0.25)\cdot 10^{-3}\,\text{ps}^{-1}\,,\nonumber\label{eq:SMpred:d:01}\\
&\left[\asls\right]_{\textrm{SM}}  = (2.0\pm 0.3)\cdot 10^{-5}\,,\nonumber\quad
&\left[\DGs\right]_{\textrm{SM}}  = (0.090\pm 0.008)\,\text{ps}^{-1} \,.\hfill \label{eq:SMpred:s:01}
\end{eqnarray}
The smallness of $\asld$ and $\asls$ can be traced back to the $(m_c/m_b)^2$ suppression in \eq{eq:G12q:03}: the expected leading contribution, proportional to $c$, is real and thus absent; the hierarchy of the CKM matrix further suppresses $\asls$. For $\aslb$, 
\begin{equation}
\aslb=\frac{\asld+g\asls}{1+g},\quad g=f\,\frac{\Gamma_d}{\Gamma_s}\frac{(1-y_s^2)^{-1}-(1+x_s^2)^{-1}}{(1-y_d^2)^{-1}-(1+x_d^2)^{-1}}\,,\quad 
y_q=\frac{\DGq}{2\Gamma_q},\,x_q=\frac{\DMBq}{\Gamma_q},\label{eq:SM:AbSL:01}
\end{equation}
and $f$ is the $B_s$--$B_d$ fragmentation fraction ratio in the $B$ sample, $f=0.269\pm 0.015$. Then,
\begin{equation}
\left[\aslb\right]_{\textrm{SM}}= (-2.40\pm 0.45)\cdot 10^{-4}\,,\label{eq:SM:AbSL:03}
\end{equation}
to be compared with the D0 result \cite{Abazov:2013uma}, $\aslb=(-4.96\pm 1.53\pm 0.72)\cdot 10^{-3}$.\\
\noindent Underlying these SM results are two important assumptions:
\begin{itemize}
\item[(i)] a single weak amplitude, the one-loop induced one with virtual top quarks, dominates $\Mmixq$,
\item[(ii)] the CKM matrix is $3\times 3$ unitary, and thus the would-be leading contribution to $\Gmixq$ has the same phase as $\Mmixq$.
\end{itemize}

\section{NP analyses: $\mathbf{3\times 3}$ unitarity and beyond \label{SEC:NP}}
\begin{figure}[h!]
\begin{center}
\subfigure[$\DC$ vs. $\asld$.\label{fig:NoUnNP:01a}]{\includegraphics[width=0.31\textwidth]{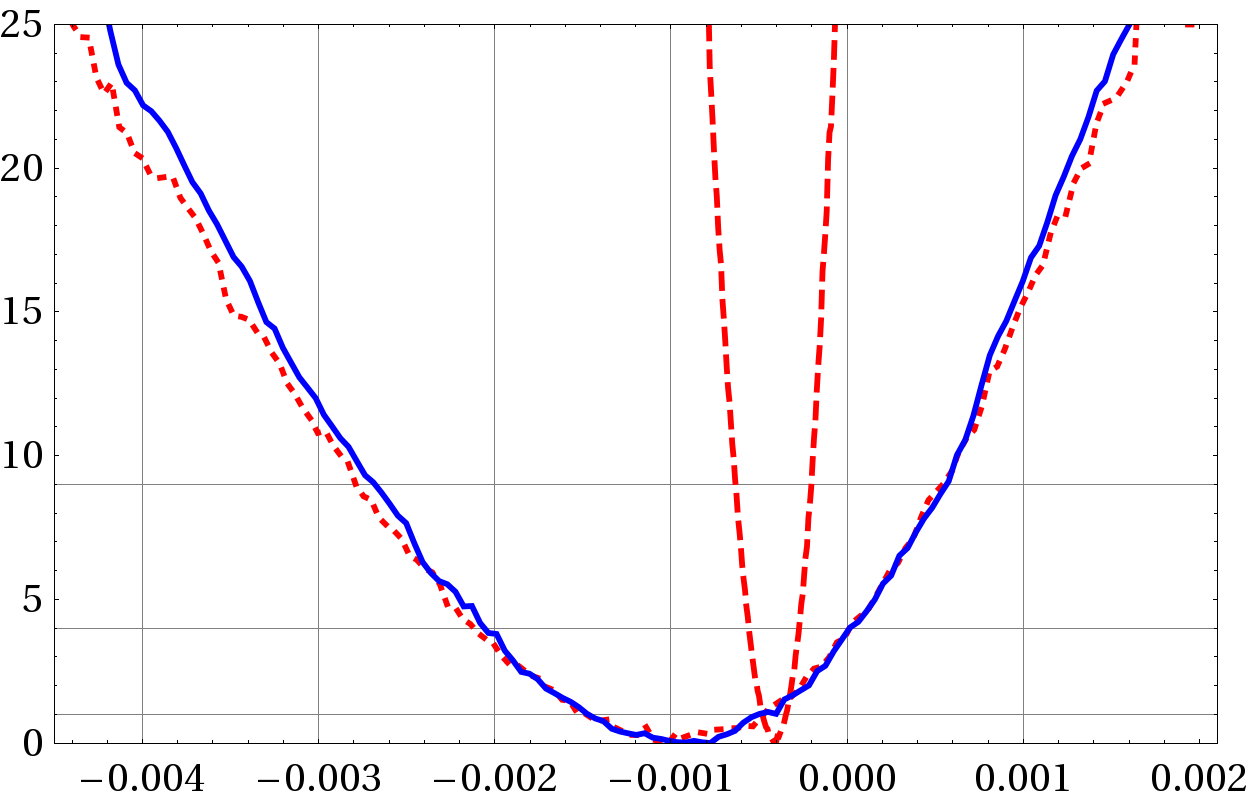}}\quad
\subfigure[$\DC$ vs. $\asls$.\label{fig:NoUnNP:01b}]{\includegraphics[width=0.3\textwidth]{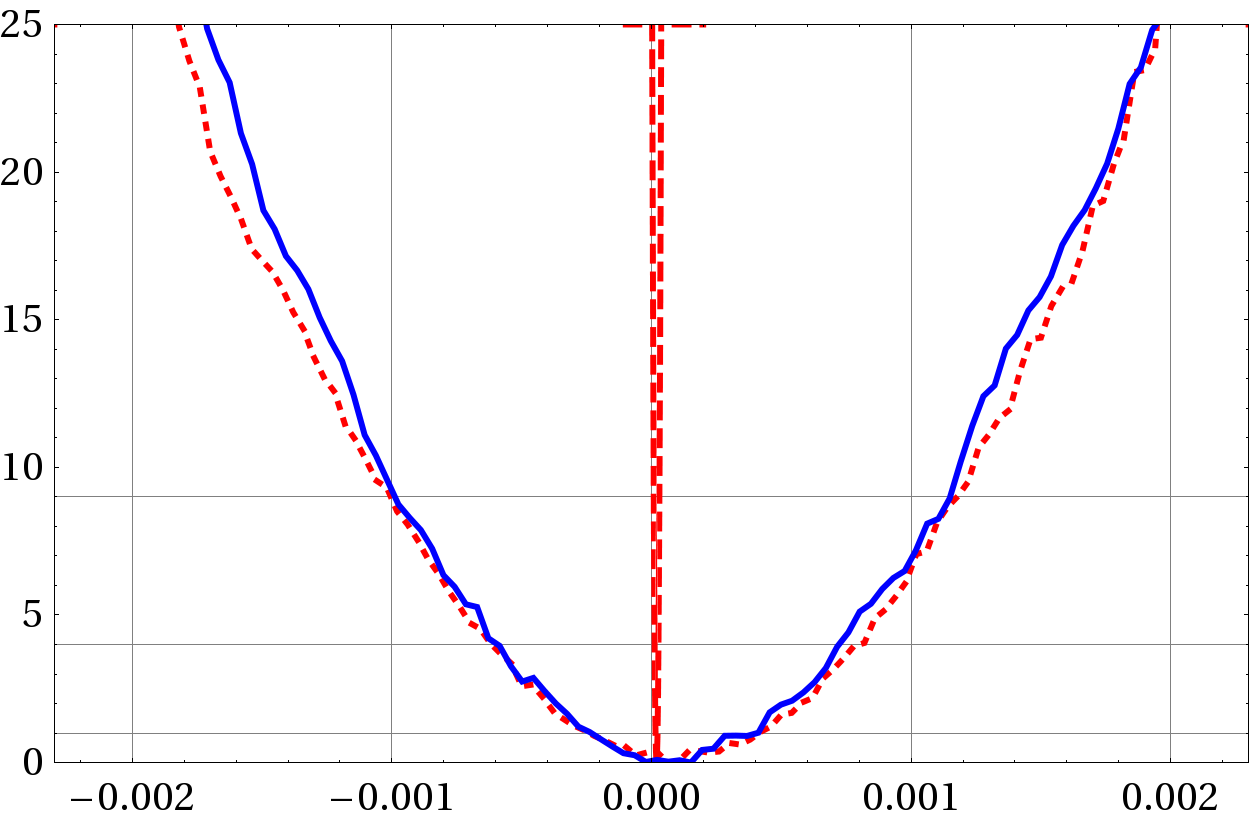}}\quad
\subfigure[$\DC$ vs. $\aslb$.\label{fig:NoUnNP:03}]{\includegraphics[width=0.3\textwidth]{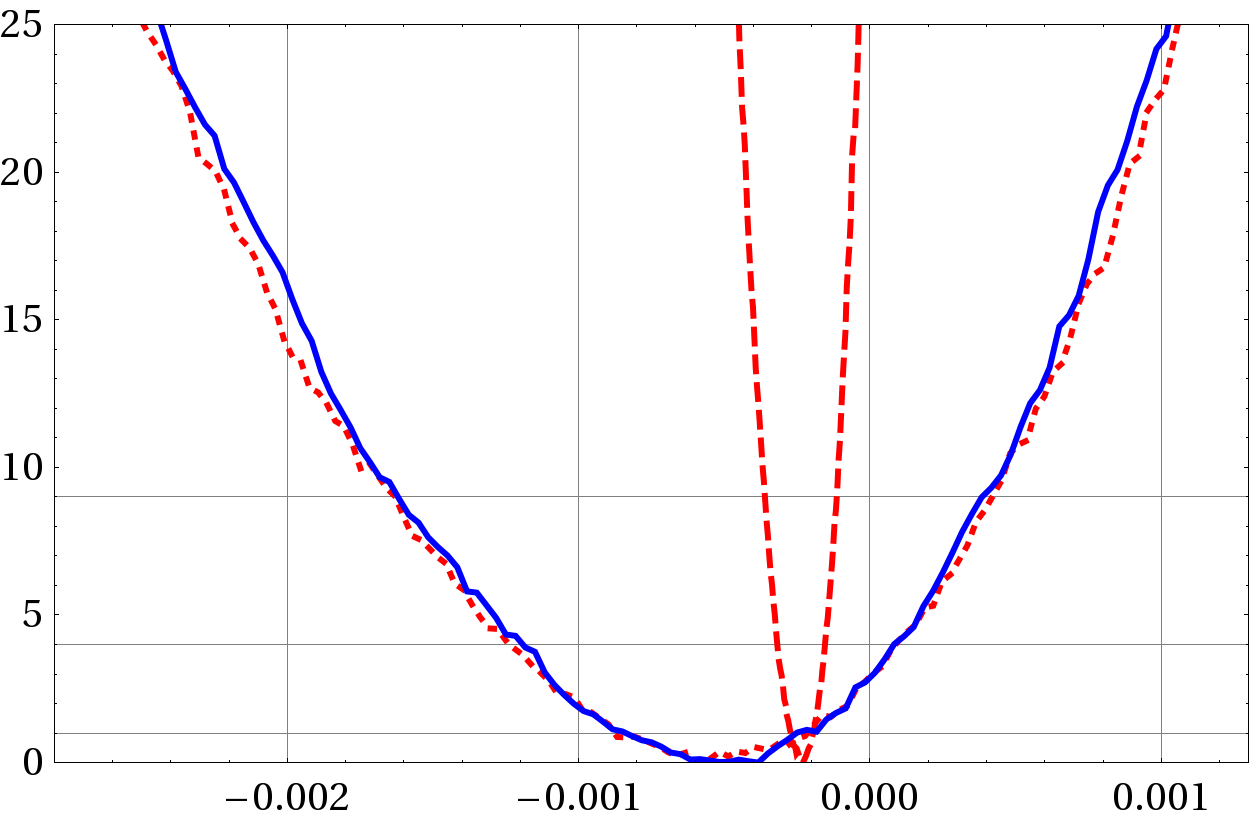}}
\caption{$\DC$ profiles of the semileptonic asymmetries from \cite{Botella:2014qya}; the {\color{blue} blue} line is the non $3\times 3$ unitary NP scenario -- eqs. \refeq{eq:NoUn:01} and \refeq{eq:NoUnMixq:01} --, the {\color{red} red dotted} line is the $3\times 3$ unitary NP scenario of \refeq{eq:33NP:Mix01} and the {\color{red} red dashed} line, the SM case. The D0 measurement is $\aslb=(-4.96\pm 1.69)\cdot 10^{-3}$ \cite{Abazov:2013uma}. \label{fig:NoUnNP:01}}
\end{center}
\end{figure}

\begin{figure}[h!]
\begin{center}
\subfigure[$\aslb$ vs. $|\V{ub}|$.\label{fig:NoUnNP:04a}]{\includegraphics[width=0.38\textwidth]{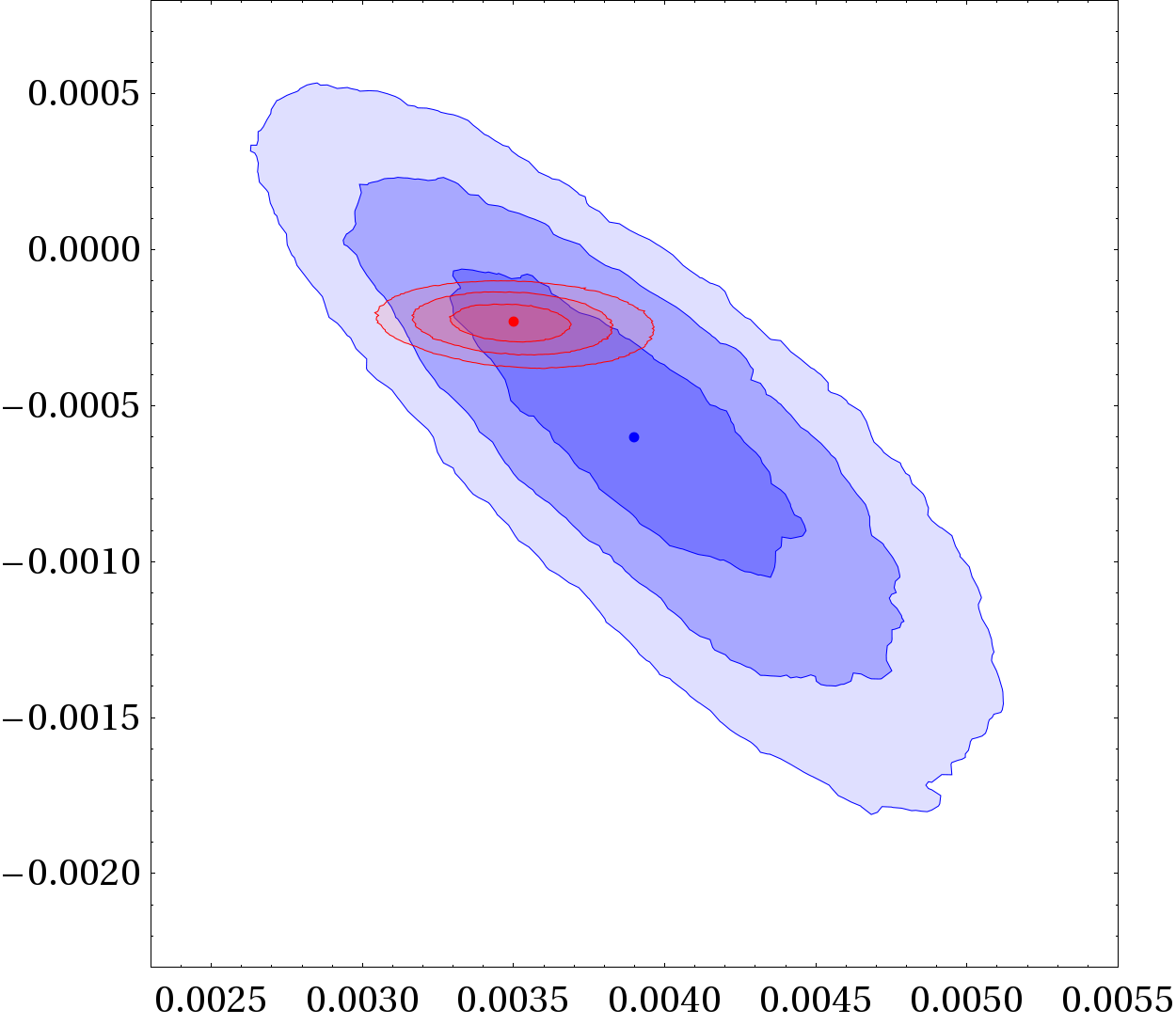}}\qquad
\subfigure[$\aslb$ vs. $\AJPP$.\label{fig:NoUnNP:04b}]{\includegraphics[width=0.36\textwidth]{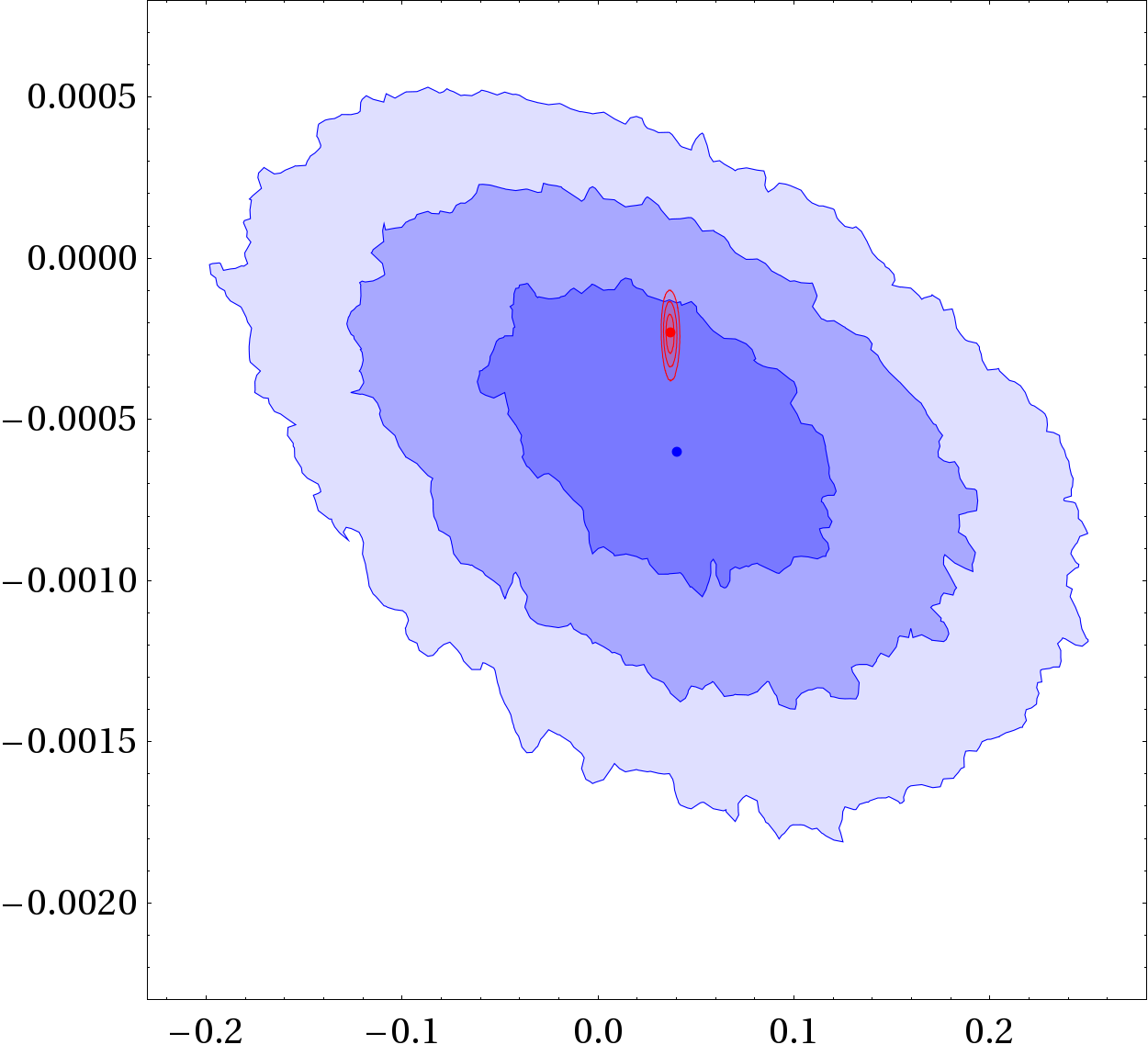}}
\caption{$\DC$ 68\%, 95\% and 99\% CL regions from \cite{Botella:2014qya}. {\color{blue} Blue} regions correspond to the non $3\times 3$ unitary NP scenario, {\color{red} red} regions correspond to the SM case.\label{fig:NoUnNP:02}}
\end{center}
\end{figure}

Since the CKM paradigm provides a consistent tree level picture of flavour changing processes, a popular class of beyond SM analyses \cite{Botella:2006va} considers that NP only affects the dispersive amplitudes $\Mmixq$ in a  manner:
\begin{equation}
\Mmixq = \left[\Mmixq\right]_{\text{SM}}\,r_q^2\,e^{-i\,2\phi_q}\ .\label{eq:33NP:Mix01}
\end{equation}
Deviations from $(r_q,\phi_d)=(1,0)$ signal the presence of NP in the mixing of $B_q$ mesons: $r_q$ modifies the SM prediction for $\DMBq=2|\Mmixq|$ while $\phi_q$ modifies mixing-induced time dependent CP asymmetries as in $B_d^0\to J/\Psi K_S$ and $B_s^0\to J/\Psi\Phi$. Since $\phi_q\neq 0$ invalidates the equality among the phases of $\Mmixq$ and the leading contribution to $\Gmixq$, the SM suppression can be lifted. \\
\noindent Nevertheless, \eq{eq:33NP:Mix01} does not exhaust the NP scenarios that could enhance the semileptonic asymmetries and thus $\aslb$: if the CKM matrix is not $3\times 3$ unitary but, on the contrary, part of a larger unitary matrix, there are additional fields beyond the standard three chiral ones. They will give new contributions to $\Mmixq$, controlled by mixings beyond the usual $3\times 3$ ones. If
\begin{equation}
\V{ub}\Vc{uq}+\V{cb}\Vc{cq}+\V{tb}\Vc{tq}\equiv-N_{bq}\neq 0\,,\label{eq:NoUn:01}
\end{equation}
modified $\Mmixq$ expressions with the following structure \cite{Barenboim:1997pf} should be considered:
\begin{equation}
\Mmixq = \frac{G_F^2M_W^2}{12\pi^2}\,M_{B_q}f_{B_q}^2B_{B_q}\eta_B
\left((\V{tb}\Vc{tq})^2S_0(x_t)+(\V{tb}\Vc{tq})\,N_{bq}\,C_1+N_{bq}^2\,C_2\right)\,.\label{eq:NoUnMixq:01}
\end{equation}
$C_1$ and $C_2$, both real, are common to both $\Mmixd$ and $\Mmixs$: all the new flavour dependence and CP violation is confined to the mixings $N_{bq}$.  Interesting examples of such scenarios are models where the fermion spectrum is extended through additional vector-like quarks \cite{Barenboim:1997pf}. For each specific model, $C_1$ and $C_2$ are then related to fundamental parameters like, e.g., masses. Equations \refeq{eq:NoUn:01} and \refeq{eq:NoUnMixq:01} can provide the central ingredient to escape the SM suppression in two ways: instead of $c$ in  \eq{eq:G12q:03}, one is left with
\begin{equation}
c\,\frac{(\La{t}{bq}+N_{bq})^2}{(\La{t}{bq})^2S_0(x_t)+2(\La{t}{bq}N_{bq}) C_{1}+(N_{bq})^2 C_{2}}\,,
\end{equation}
which, in general, is not real, and thus the semileptonic asymmetries could be enhanced \cite{Botella:2014qya}. One can thus conduct NP ``model independent'' analyses similar to the ones considering \eq{eq:33NP:Mix01}, now including beyond $3\times 3$ unitarity. Figure \ref{fig:NoUnNP:01} shows the resulting $\DC$ profile for the semileptonic asymmetries in this kind of NP scenarios together with the profiles corresponding to scenarios where $3\times 3$ unitarity is maintained but NP introduced in $\Mmixq$ -- \eq{eq:33NP:Mix01} --, and to the SM case. In both NP scenarios, semileptonic asymmetries can reach values at the $10^{-3}$ level. Despite the very significant increase with respect to the SM case, those values are still insufficient to account for the D0 measurement.
It is important to stress that those enhancements are correlated to deviations in some related observables: in the $bd$ sector, reaching $\asld$ values at the $10^{-3}$ level requires $|\V{ub}|$ to deviate from the tight SM $|\V{ub}|$--$\AJPKs$ connection; in the $bs$ sector, reaching $\asls$ values at the $10^{-3}$ level requires $\AJPP$ to depart from the SM expectation $\AJPP\simeq 0.04$. Both ingredients can be present in NP scenarios with the CKM matrix part of a larger unitary matrix. The same-sign dimuon asymmetry $\aslb$ inherits such correlated deviations, as figures \ref{fig:NoUnNP:04a} and \ref{fig:NoUnNP:04b} show.

\section{Conclusions}
Scenarios where the CKM mixing matrix is not $3\times 3$ unitary but part of a larger mixing matrix, as e.g. SM extensions including vector-like quarks, provide incorporate the ingredients that can avoid the SM suppression of semileptonic asymmetries in $B$ meson systems and enhance the same-sign dimuon asymmetry at the $10^{-3}$ level. NP scenarios with $3\times 3$ unitarity and arbitrary contributions in the $B$ mixings can accommodate similar enhancements. Despite these significant departures from the SM, the values remain insufficient to reproduce the D0 measurement.


\end{document}